# Photonics-based short-time Fourier transform without high-frequency electronic devices and equipment

Pengcheng Zuo[a,b], Dong Ma[a,b], and Yang Chen[a,b,*]

[a] *Shanghai Key Laboratory of Multidimensional Information Processing, East China Normal University, Shanghai 200241, China*
[b] *Engineering Center of SHMEC for Space Information and GNSS, East China Normal University, Shanghai 200241, China*
[*] ychen@ce.ecnu.edu.cn

**ABSTRACT**
A photonics-based short-time Fourier transform (STFT) system is proposed and experimentally demonstrated based on stimulated Brillouin scattering (SBS) without using high-frequency electronic devices and equipment. The wavelength of a distributed feedback laser diode is periodically swept by using a low-speed periodic sawtooth/triangular driving current. The periodic frequency-sweep optical signal is modulated by the signal under test (SUT) and then injected into a section of SBS medium. The optical signal from another laser diode as the pump wave is reversely injected into the SBS medium. After simply detecting the forward transmission optical signals in a low-speed photodetector, the STFT of the SUT can be implemented. The system is characterized by the absence of any high-frequency electronic devices or equipment. An experiment is performed. The STFT of a variety of RF signals is carried out in a 4-GHz bandwidth. The dynamic frequency resolution is demonstrated to be around 60 MHz.

**Key word:** short-time Fourier transform, stimulated Brillouin scattering, time-frequency analysis, distributed feedback laser diode.

## 1. Introduction

Short-time Fourier transform (STFT) is one of the typical time-frequency analysis methods, which is conventionally implemented in the electrical domain. STFT is widely used in the fields of speech, image processing, and radar signal processing [1]-[4]. However, limited by the sampling rate of the analog-to-digital converter, the processing speed of STFT based on digital signal processing in the electrical domain is relatively slow, resulting in that the analysis bandwidth and frequency are inherently limited [5], [6].

Over the past two decades, there has been a considerable interest in using photonic devices to implement flexible microwave measurement functions [7]-[9], which has been proved to be able to overcome the bottleneck of electrical solutions with the assistance of modern photonics [10]. However, few photonics-based attempts have been proposed to provide complete time-frequency information due to the lack of temporal dimension information. Only a few photonic solutions for implementing STFT based on dispersive devices have been proposed [11]-[13]. Generally, dispersion-based STFT has two main drawbacks: 1) high-frequency resolution is highly dependent on a large amount of dispersion; 2) the dispersive devices are hard to be reconstructed in real-time in multiple dimensions, which constrains the reconfigurability of STFT. The proposal in [13] reduces the dispersion values required to implement a high-frequency resolution, but it still suffers from the above second restriction. In addition, the analysis bandwidths in [12] and [13] are limited to no more than 2.5 GHz. Further extending the analysis bandwidth may place higher requirements on the implementation and complexity of the system.

To avoid the limitations of the existing dispersion-based STFT, most recently, we have proposed the stimulated Brillouin scattering (SBS)-based STFT [14]. The electrical signal under test (SUT) is converted to the optical domain and temporally segmented by using a periodic frequency-sweep optical signal. In each sweep period, the SUT is considered stationary, and the corresponding frequency information is obtained by using the SBS-based frequency-to-time mapping (FTTM). The whole process is equivalent to implementing the STFT of the SUT. A dynamic frequency resolution of 60 MHz and an observation bandwidth of 12 GHz are achieved and only limited by the equipment used in the experiment. The major drawback of the scheme in [14] is that a high-speed electrical arbitrary waveform generator (AWG) is needed to generate the periodic frequency-sweep optical signal, which increases the cost and complexity of the system. In this Letter, a photonics-based STFT system is proposed and experimentally demonstrated without using high-frequency electronic devices and equipment. An experiment is carried out to verify the feasibility and effectiveness of the system. The STFT of a variety of RF signals is successfully realized in a 4-GHz bandwidth, and the dynamic frequency resolution is also demonstrated to be around 60 MHz.

## 2. Principle and Experimental Setup

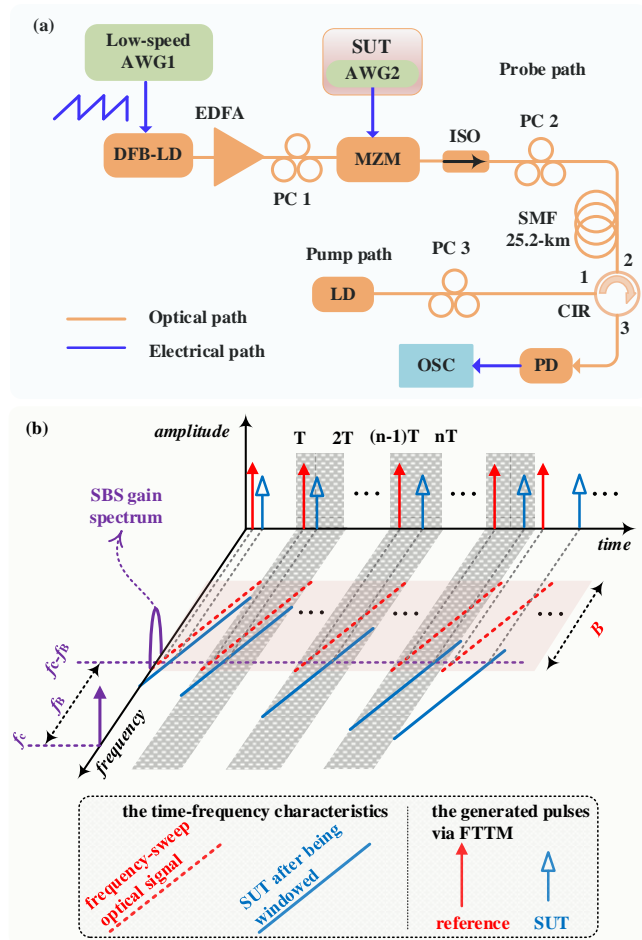

Fig. 1. (a) Schematic diagram of the proposed SBS-based real-time all-optical STFT system. (b) Operation principle of the SBS-based STFT. LD, laser diode; DFB-LD, distributed feedback laser diode; MZM, Mach–Zehnder modulator; AWG, arbitrary waveform generator; SUT, signal under test; PC, polarization controller; EDFA, erbium-doped fiber amplifier; ISO, isolator; CIR, circulator; PD, photodetector; OSC, oscilloscope.

Fig. 1(a) schematically shows the proposed SBS-based STFT system, which consists of a typical SBS pump-probe setup. In the pump path, a continuous-wave (CW) light wave centered at $f_c$ from a laser diode (LD, ID Photonics CoBriteDX1-1-C-H01-FA) is directly sent to an optical circulator and then launched into the SBS medium, i.e., a section of 25.2-km single-mode fiber (SMF) in this experiment, where it interacts with the counter-propagating probe wave from the probe path. An SBS gain spectrum with its frequency centered at $f_c$-$f_B$ is thus generated as shown in Fig. 1(b). Here, $f_B$ is the Brillouin frequency shift. In the probe path, the wavelength of a distributed feedback laser diode (DFB-LD, Lucent D2511G) is periodically swept by applying a low-speed periodic sawtooth current from a low-speed AWG (AWG1, RIGOL DG2052) to generate a periodic frequency-sweep optical signal. The time-frequency characteristic of the frequency-sweep optical signal from the DFB-LD is shown in the red dotted line in Fig. 1(b). The sweep period $T$, bandwidth $B$, and frequency-sweep range can be changed by adjusting the period and amplitude of the sawtooth current and the center frequency of the DFB-LD, respectively. Subsequently, the frequency-sweep optical signal, as a new carrier, is amplified by an erbium-doped fiber amplifier (EDFA, Amonics EDFA-PA-35-B), and then modulated at a Mach–Zehnder modulator (MZM, Fujitsu FTM7938EZ) by the SUT from AWG2 (Keysight M8190A). Note that the bias point of the MZM is set slightly off the null point to make the carrier and the first-order optical sidebands have close power. In Fig. 1(b), the SUT is chosen as a linearly frequency-modulated (LFM) signal for example. It should be noted that the negative sideband that does not interact with the SBS gain is not shown in the figure. Then, the output of the MZM is injected into the 25.2-km SMF through an optical isolator. Polarization controller 1 (PC1) is used to optimize the light polarization before the MZM. PC2 and PC3 are used to ensure the most efficient stimulated Brillouin interaction. After SBS interaction with the pump wave from the pump path, the optical signal from the circulator is detected in a photodetector (PD, Nortel PP-10G) and monitored by a low-speed oscilloscope (OSC, RIGOL DS1104Z Plus).

As the sweep period $T$ is much smaller than the time duration of the SUT, the SUT is considered to be approximately stationary over a sweep period, which is the basic assumption and premise of STFT. Therefore, in each sweep period $T$, the frequency components of the SUT after being windowed are mapped to the time domain via SBS-based FTTM in the format of low-speed electrical pulses as shown in Fig. 1(b). Pulses in different periods represent the frequency components of the SUT in the corresponding time windows, which are then processed and recombined to obtain the time-frequency diagram of the SUT. Compared with the system in [14], the reference signal is not used in the proposed system because the unsuppressed optical carrier can be used to provide a label of time and frequency during the recombination process.

## 3. Experimental results and discussion

In the experiment, a section of 25.2-km SMF is used as the SBS medium, and the corresponding Brillouin frequency shift $f_B$ is around 10.8 GHz. Compared with the method using a high-speed frequency-sweep electrical signal to realize optical wavelength sweeping [14], although the use of a sawtooth current to drive the DFB-LD makes the system have a simpler structure, there will be undesirable frequency sweep characteristics near the jump point of the sawtooth current. To reduce the difficulty of analyzing and processing the electrical pulses, we use a triangular wave as the driving signal in the experiment. In this case, the first half period and the second half period of each sweep period contain the same information, and only half of the obtained pulses are used to realize the STFT.

To preliminarily verify the time-frequency analysis function of the proposed STFT system, the triangular voltage with a peak-to-peak amplitude of 600 mV is applied to the DFB-LD. The DC offset current of the DFB-LD is set to 40 mA and the temperature is set to 23.860 °C by the LD controller (Thorlabs ITC4001). The frequency of the triangular wave is set to 100 kHz, which is equivalent to a sweep period of 10 μs. In this case, the bandwidth of the usable frequency-sweep optical signal is about 2.7 GHz. The SUT is an LFM signal with a bandwidth ranging from 0.2 GHz to 2.2 GHz and a time duration of 500 μs or 1500 μs. Figs. 2 (a) and (b) show the measured time-frequency diagrams of the LFM signal using the proposed STFT system. Under different signal lengths, the time-frequency characteristics of the signal are well measured. It is noted that the time-frequency diagram obtained is not an ideal linear curve, which is caused by the fact that the frequency-sweep optical signal generated by the triangular wave-driven DFB-LD still has a certain frequency-sweep nonlinearity. The sweep nonlinearity can be improved by using a better laser or nonlinear compensation. The time resolution can be easily changed by adjusting the frequency of the driving signal. Then, the frequency is set to 50 kHz, which corresponds to a sweep period of 20 μs. Figs. 2 (c) and (d) show the measured time-frequency diagrams. Compared with Figs. 2 (a) and (b), it can be concluded that when the sweep period is too long, the time resolution will be poor, and obvious error will be caused because local stationarity is not satisfied. In the experiment, the sweep period of the driving signal is not further decreased due to the limitation of the LD controller (Thorlabs ITC4001).

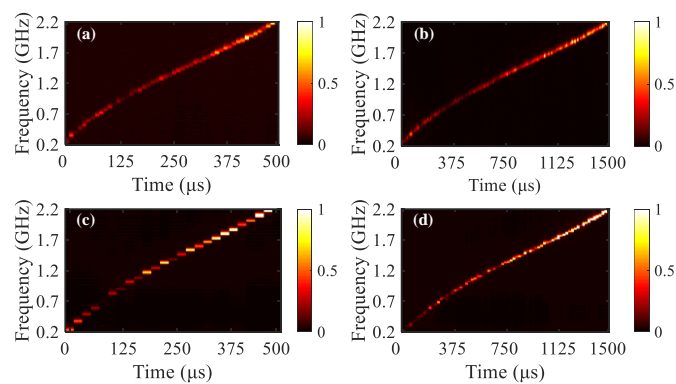

Fig. 2. Measured time-frequency diagrams of the LFM signals with a bandwidth ranging from 0.2 GHz to 2.2 GHz and a time duration of 500 μs and 1500 μs. The frequency of the triangular wave is (a) and (b) 100 kHz, (c) and (d) 50 kHz.

The capability of analyzing different format signals is then demonstrated. The system parameters are kept unchanged except that the frequency of the driving signal is set to 100 kHz. In this case, the sweep period is 10 μs. Four kinds of broadband signals with bandwidth from 0.2 GHz to 2.2 GHz are chosen as the SUTs and the time duration of the signals are all set to 1500 μs. The time-frequency analysis results for these four kinds of signals are shown in Figs. 3 (a)-(d). Fig. 3(e) shows the measured time-frequency diagram for a 1680-μs specially designed signal whose time-frequency characteristic is the abbreviation "ECNU" of "East China Normal University". As can be seen, the time-frequency diagrams of all these kinds of signals are well constructed with good resolution.

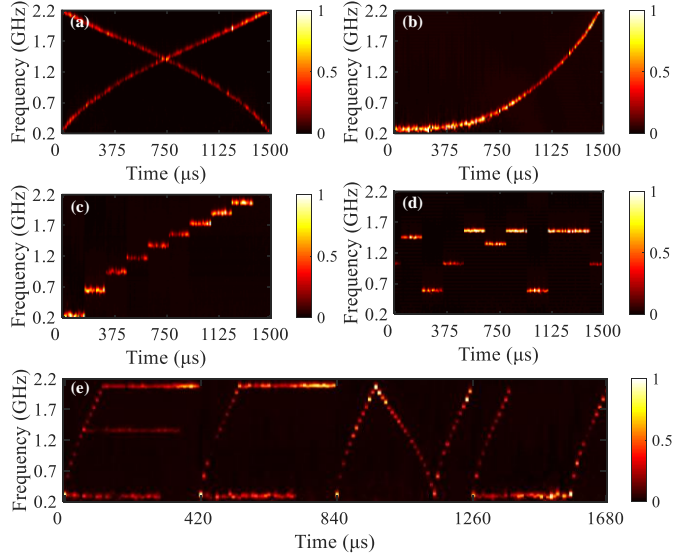

Fig. 3. Measured time-frequency diagrams by the proposed STFT scheme for (a) dual-chirp LFM signal, (b) non-linearly frequency-modulated signal, (c) step-frequency signal, (d) frequency-hopping signal, (e) a specially designed signal whose time-frequency characteristic is the abbreviation "ECNU" of "East China Normal University".

The dynamic frequency resolution of the proposed STFT systems is further demonstrated using a two-tone signal as the SUT. The system parameters are completely consistent with those used to obtain the results in Fig. 3. Figs. 4(a) and (b) show the measured time-frequency diagrams of the two-tone signal with frequency intervals of 70 MHz and 60 MHz, respectively. It is observed that the frequency resolution is around 60 MHz. The frequency resolution is consistent with the analysis and measurement results in [14] when the sweep rate is close to that of this work.

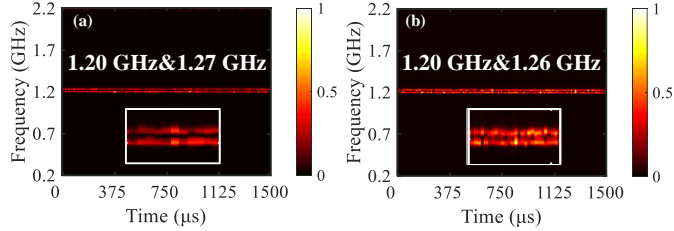

Fig. 4. Measured time-frequency diagrams by the proposed STFT scheme for a two-tone signal with frequencies of (a) 1.20 GHz and 1.27 GHz, (b) 1.20 GHz and 1.26 GHz.

In the above experiment, the analysis bandwidth of the system is 2.2 GHz, which is determined by the linear frequency sweep range of DFB-LD driven by the time-varying driving signal. To demonstrate the analysis bandwidth of the STFT system on a wider bandwidth, another DFB-LD (DFB-1552-40-SM) is used. In this case, the triangular wave driving signal with a peak-to-peak amplitude of 1.5 V is applied to the DFB-LD. The DC offset current of the DFB-LD is set to 100 mA, and the temperature is set to 23.860 °C. The sweep period of the driving signal is also set to 10 μs. Using this DFB-LD and the above system parameters, the bandwidth of the frequency-sweep optical signal is about 6.5 GHz. Signals of different formats from 0 to 4 GHz are chosen as SUTs. The analysis results are shown in Fig.5. The time-frequency diagrams of different kinds of signals are all successfully constructed. By comparing the time-frequency diagrams of LFM signals in Fig. 3(a) and Fig. 5(a), it can be seen that the time-frequency diagram recovered in this experiment has better linearity, indicating that the DFB-LD used in this experiment has better sweep linearity when driven by the linear time-varying current. However, because the stability of this DFB-LD is worse, the obtained time-frequency diagrams in Fig. 5 have obvious jitters and its frequency resolution is not as good as that shown in Fig. 3 and in [14] under a similar sweep rate. In practical applications, the stability of the STFT system can be improved by using a more stable laser and by correlating two lasers.

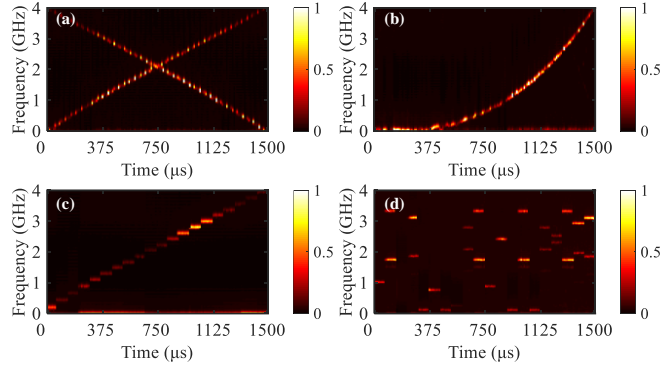

Fig. 5. Measured time-frequency diagrams by the proposed STFT scheme for (a) dual-chirp LFM signal, (b) non-linearly frequency-modulated signal, (c) step-frequency signal, (d) frequency-hopping signal, with a bandwidth ranging from 0 to 4 GHz.

Compared with the dispersion-based all-optical STFT schemes [11]-[13], the proposed system has better reconfigurability by directly manipulating the DFB-LD. Although we only show the operating bandwidth of 2.2 GHz and 4 GHz in this work, the operating bandwidth of the proposed method can be greatly widened by selecting the appropriate DFB-LD. Compared with other microwave time-frequency analysis methods based on SBS [15], [16], the major improvement is that the proposed SBS-based STFT is real-time because signal receiving and instantaneous acquisition of the time-frequency information are performed at the same time.

Compared with our previous work in [14], the main advantage of the system is that the structure and cost are greatly reduced without using high-frequency electronic devices and equipment: The high-speed AWG used to generate a high-speed frequency-sweep electrical signal for the generation of a wavelength-sweep optical signal in [14] is replaced by a simple DFB-LD, which only requires a relatively low-frequency time-varying driving signal. Furthermore, the additional electrical reference signal in [14] is not employed in the proposed system because the unsuppressed carrier is used as the reference, which also simplifies the implementation of the STFT system.

Due to the lack of a highly nonlinear fiber in our laboratory, a section of 25.2-km SMF is used as the SBS medium. The STFT in this work is real-time when the transmission delay in the SBS medium, i.e., the SMF, is not considered. When the SBS medium is taken into account, the real-time performance of the proposed system can be further improved by using a short high-gain SBS medium [17] or a chip-based SBS medium [18].

## 4. Conclusions

In summary, we have optimized the design of the SBS-based STFT and experimentally demonstrated a simple and low-cost all-optical STFT scheme without using high-frequency electronic devices and equipment. The feasibility and effectiveness of the proposed system have been verified by an experiment. STFTs of different kinds of RF signals are carried out. The dynamic frequency resolution and analysis bandwidth are demonstrated to be 60 MHz and 4 GHz, respectively. The proposed system has avoided the drawbacks of the existing dispersion-based STFTs and thus provides a promising alternative for microwave time-frequency information measurement. Benefiting from integrated photonics and chip-based SBS, we also believe that the proposed SBS-based STFT can be better used in real-world applications, such as electromagnetic spectrum monitoring in electronic warfare.

**Funding**

National Natural Science Foundation of China (NSFC) (61971193); Natural Science Foundation of Shanghai (20ZR1416100); Science and Technology Commission of Shanghai Municipality (18DZ2270800).

**Conflicts of interest**

The authors declare no conflicts of interest.